# State-of-charge estimation of lithium-ion batteries using a tree seed and genetic algorithm-optimized generalized mixture minimum error entropy-based square root cubature Kalman filter

* Haiquan Zhao[a,b], Xiong Yin[a,b], Jinhui Hu[a,b]

[a] *School of Electrical Engineering, Southwest Jiaotong University, Chengdu 610031, China*

[b] *Key Laboratory of Magnetic Suspension Technology and Maglev Vehicle, Ministry of Education*

**Abstract:**

The cubature Kalman filter based on minimum error entropy (MEE-CKF) offers accurate and robust performance in state of charge (SOC) estimation. However, due to the inflexibility of the minimum error entropy (MEE), this algorithm demonstrates limited robustness when confronted with more complex noise environments. To address these limitations, this paper proposes a generalized mixture minimum error entropy-based (GMMEE) square-root cubature Kalman filter (GMMEE-SRCKF). The square-root algorithm ensures improved numerical stability and avoids covariance degeneration, while the GMMEE criterion with two flexible kernels adapts effectively to non-Gaussian noise. Moreover, a hybrid tree seed and genetic algorithm (TSGA) is introduced to optimize the kernel parameters automatically. Experimental results confirm that the TSGA-optimized GMMEE-SRCKF outperforms existing robust filters, achieving the root mean square error (RMSE) of less than 0.5%.

## 1 Introduction

As environmental pollution and energy scarcity challenges become more pressing, renewable energy sources with zero emissions are gradually replacing traditional energy sources [1]. Due to its

This work was supported by the National Natural Science Foundation of China (grant: 62171388, 61871461, 61571374). (*Corresponding author:Haiquan Zhao).

high energy density, eco-friendliness, and broad range of working temperatures, the lithium-ion battery is a commonly used energy storage device [2]. Nevertheless, the battery management systems (BMS) need to guarantee an optimal allocation of battery energy to maintain the long-term safety and efficiency of the system [3]. Estimating the state of charge (SOC) accurately is vital for the battery system as it helps prevent overcharging and enhances performance [4,5]. However, due to the high degree of non-linearity within the cell, direct measurements can not provide accurate SOC values [6]. Instead, voltage, current, and temperature data must be employed for indirect estimation [7,8].

Nowadays, the SOC estimation methods are primarily categorized into three types [9,10]: traditional state estimation [11,12], data-driven (DD) methods [13-16], and model-driven (MD) methods [17,18]. The disadvantage of the open-circuit voltage method in traditional state estimation is that it takes a considerable period before the battery reaches a steady state, limiting its practicality in SOC estimation [19]. While DD approaches do not require detailed battery models, they rely on large amounts of battery data and significant computational resources to map the relationship between SOC and models [7,20].

Compared to the previously described algorithms, the MD methods incorporate an accurate battery model and reasonable filtering algorithms to achieve effective SOC estimation. This approach maintains high estimation accuracy while attaining high computational efficiency [21]. The most commonly utilized battery models include the electrochemical models (EM) [17] and the equivalent circuit models (ECM) [18,22,23]. However, the EM requires substantial electrochemistry knowledge and the ability to solve partial differential equations, imposing a significant burden on designers and limiting its practical applicability in BMS. To address this challenge, many studies have highlighted the use of ECM [24,25]. ECM primarily employs electrical components to simulate lithium batteries, with second-order resistor–capacitor (RC) circuits being particularly prevalent due to their simple structure and good real-time performance [26]. In addition, the parameter identification is another important issue in SOC estimation, which can be categorized into online [27] and offline [28,29] methods. Although online parameter identification allows capturing the dynamic changes of the battery, it is often challenging to accurately track parameter variations in practical SOC estimation [26]. The offline method is a prevalent approach for identifying battery parameters, largely due to its simplicity and accuracy [30].

Once an accurate equivalent model has been selected, the crucial issue is how to carry out an accurate SOC estimation with a robust algorithm. The Kalman filter (KF) and its variants are particularly effective among SOC estimation algorithms [31,32]. The extended Kalman filter (EKF) linearizes the nonlinear measurement function of the ECM via Taylor series expansion, but its performance is highly sensitive to the initial error covariance [33]. To address this limitation, the unscented Kalman filter (UKF) has been proposed, which improves both accuracy and robustness compared with the EKF [34,35]. In the UKF, the unequal weighting of 2n+1 sigma points, with the central point assigned the largest weight, can result in significant errors and even cause filter divergence in nonlinear estimation. Cubature Kalman filter (CKF) is another improved approach, as it can achieve third-order accuracy and higher numerical stability. In addition, CKF-based SOC estimation is more efficient than UKF, as it employs fewer sampling points than UKF [36]. The results in [37] demonstrated that the CKF-based algorithms exhibit superior accuracy. Furthermore, to address asymmetric and non-positive definite covariance during iteration, Peng [26] applied the square root (SR) concept to the cubature Kalman filter (SRCKF), effectively avoiding the matrix-squaring operation and achieving superior results. In [8], a hybrid approach integrating DD and MD methods is proposed to enhance SOC estimation using SRCKF, which can accurately track the decreasing trend of SOC. However, the abovementioned methods are grounded in the minimum mean square error (MMSE) criterion. The performance of MMSE-based Kalman filters (KFs) in SOC estimation tends to decline under complex non-Gaussian noise [38].

To address the issues mentioned above, Chen [38] employed the maximum correlation criterion (MCC) and integrated it with the Kalman filter (MCC-KF). MCC-based CKF has been adopted for target-tracking applications [39,40]. The findings [6,7] indicated that these filters enhance accuracy under non-Gaussian noise. Although MCC demonstrates robust performance in handling non-Gaussian noise, its efficacy may be constrained when confronted with more complex multimodal non-Gaussian noise [41]. Furthermore, [42] presented the SRCKF based on the minimum error entropy (MEE) criterion in developing the SOC estimation, proving it to be more robust than the conventional SRCKF. However, MEE filters employing a single error entropy limit their application flexibility and fail to address more complex non-Gaussian noise. In [43], a mixture minimum error entropy (MMEE) cubature Kalman filter (MMEE-CKF) is proposed. A comparative analysis demonstrates that the UKF based on the MMEE criterion is more accurate for target tracking [44]. Furthermore, the MMEE criterion-based KFs typically employ Gaussian functions as the kernel functions, which cannot change their shape flexibly to adapt to non-Gaussian noise. Wen [45] proposed a generalized maximum correlation criterion (GMCC), which employs the generalized Gaussian function to estimate SOC under complex non-Gaussian noises accurately.

Furthermore, a hybrid tree seed and genetic algorithm (TSGA) is developed to optimize the kernel parameters automatically. The tree seed algorithm (TSA) is a metaheuristic optimization method characterized by its strong local search capability, fast convergence speed, and simple algorithmic structure, making it widely applicable in various optimization tasks [46]. However, the TSA still has two critical limitations. Firstly, it tends to lose population diversity during the search process. Secondly, it is subject to convergence to local optima, particularly in complex, multimodal search spaces [47]. In contrast, the genetic algorithm (GA), another widely used metaheuristic method, exhibits superior global search capabilities through genetic operations such as crossover and mutation [48]. By hybridizing TSA with GA, it is possible to leverage the complementary strengths of both algorithms, combining the fast convergence of TSA with the enhanced global exploration ability achieved through the diversity mechanisms of GA.

This article addresses the limitations of the existing cubature Kalman filter based on minimum error entropy (MEE-CKF) for SOC estimation under complex non-Gaussian noise conditions. To improve adaptability and robustness, a new cost function is constructed by combining two MEE kernels with different bandwidths through a mixing factor and replacing the Gaussian kernels with generalized Gaussian kernels. The SRCKF is employed to enhance numerical stability and prevent covariance degeneration. In addition, a hybrid TSGA method is integrated to optimize the kernel parameters, enabling adaptive adjustment to varying noise characteristics and further improving estimation accuracy. The main contributions are as follows:

(1) To estimate the SOC under complex non-Gaussian noises, a novel algorithm combining the generalized mixture minimum error entropy (GMMEE) criterion and the square root cubature Kalman filter (GMMEE-SRCKF) is proposed.

(2) Furthermore, the MMEE criterion and the generalized minimum error entropy (GMEE) criterion can be derived from the GMMEE criterion by adjusting the parameters. The derived criteria-based CKF demonstrates superior tracking accuracy for the SOC estimation under varying operating conditions.

(3) A hybrid optimization algorithm based on the TSGA is developed to automatically optimize the kernel parameters of the GMMEE, further improving the estimation accuracy and robustness of the GMMEE-SRCKF.

The organizational framework of the paper is systematically structured across five sections. In Section 2, the second-order RC ECM and the GMMEE criterion are described. In Section 3, the TSGA integrated with the GMMEE-SRCKF is methodically developed. In Section 4, a detailed evaluation of experimental validations is provided, followed by conclusive insights in the final section.

## 2 Battery modeling and GMMEE criterion

### *2.1 Battery modeling*

The internal structure of lithium-ion batteries is highly sensitive to external factors such as temperature, and the internal electrochemical reactions are inherently complex. Although higher-order RC networks introduce more parameters, the increased model complexity may limit their fitting capability and substantially raise computational burden. In contrast, the second-order RC ECM extends the first-order RC ECM by adding an RC branch, which not only accounts for the nonlinear characteristics of the battery but also provides a more accurate representation of polarization effects during charge–discharge processes [49-51]. At the same time, its relatively low complexity ensures feasibility for practical engineering applications. Therefore, a second-order RC equivalent circuit is adopted as the ECM for lithium-ion batteries in this paper. The structure of this model is illustrated in Fig. 1.

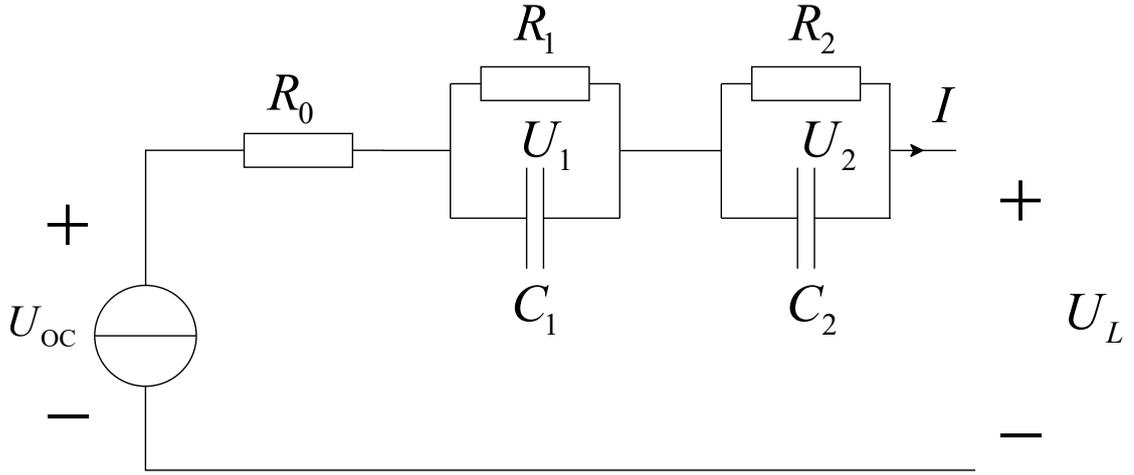

**Fig. 1** Parametric of second-order RC topology.

As the figure shows, $U_L$ means the terminal voltage; $U_{OC}$ represents the open circuit voltage; $I$ is the current; $R_0$ stands for ohmic internal resistance. Regarding polarization parameters, $R_1$ and $C_1$ characterize the electrochemical polarization RC pair, whereas $R_2$ and $C_2$ describe the concentration polarization components; $U_1$ and $U_2$ represent polarization voltage.

Based on Kirchhoff's laws, the battery's dynamic behavior can be formulated as:

$$\begin{cases} \dfrac{dU_1}{dt} = \dfrac{I(t)}{C_1} - \dfrac{U_1}{R_1 C_1} \\ \dfrac{dU_2}{dt} = \dfrac{I(t)}{C_2} - \dfrac{U_2}{R_2 C_2} \\ U_L = U_{OC}(t) - I(t)R_0 - U_1 - U_2 \end{cases} \quad (1)$$

where $I(t)$ denotes working current at time $t$, which discharge direction serves as the reference. the $U_{OC}(t)$ is the open-circuit voltage (OCV) at time $t$. The $U_{OC}(t)$ primarily involves two variables: ambient temperature and the SOC. This indicates that at a specific temperature, the OCV can be directly correlated with the SOC [52].

The SOC is calculated by the ampere-hour integration method:

$$SOC(t) = SOC(t_0) - \frac{\int \lambda I(t)dt}{Q_{max}} \tag{2}$$

where $Q_{max}$ denotes the nominal capacity, $SOC(t)$ is the SOC at time $t$, and $\lambda$ is the coulomb efficiency.

To construct discrete state space expressions that can be processed by the computer, the selected continuous-time state equations must be discretized. The system state function and measurement function will subsequently be presented as follows:

$$\begin{bmatrix} SOC_{\tau+1} \\ U_{1,\tau+1} \\ U_{2,\tau+1} \end{bmatrix} = \begin{bmatrix} 1 & 0 & 0 \\ 0 & e^{-\frac{\Delta t}{R_1 C_1}} & 0 \\ 0 & 0 & e^{-\frac{\Delta t}{R_2 C_2}} \end{bmatrix} \begin{bmatrix} SOC_\tau \\ U_{1,\tau} \\ U_{2,\tau} \end{bmatrix} + \begin{bmatrix} -\frac{\lambda \Delta t}{Q_{max}} \\ R_1 \left(1 - e^{-\frac{\Delta t}{R_1 C_1}}\right) \\ R_2 \left(1 - e^{-\frac{\Delta t}{R_2 C_2}}\right) \end{bmatrix} I_\tau + w_\tau \tag{3}$$

$$U_{L,\tau} = Uoc(SOC_\tau) - I_\tau R_0 - U_{1,\tau} - U_{2,\tau} + r_\tau \tag{4}$$

where $\tau$ is the discrete time, $r_\tau$ and $w_\tau$ are the measurement and process noise at time $\tau$, and their corresponding covariance matrices are $\mathbf{R}_\tau$ and $\mathbf{Q}_\tau$, and $\Delta t$ is the sample duration, respectively.

## 2.2 Generalized mixture minimum error entropy criterion

As a method of analyzing the degree of similarity between measurement $Y$ and estimation $X$, the information $e = Y - X$ between them can be expressed as:

$$H_\alpha(e) = \frac{1}{1-\alpha} \log V_\alpha(e) \tag{5}$$

where $\alpha\,(\alpha \neq 1,\ \alpha > 0)$ denotes the order of Rényi's entropy, and $V_\alpha(e)$ is mathematically expressed as:

$$V_\alpha(e) = \int p^\alpha(x)dx = E[p^{\alpha-1}(e)] \tag{6}$$

In this Eq., $p(\cdot)$ is the probability density function, calculated as:

$$p(x) = \frac{1}{N} \sum_{i=1}^{N} G_{\alpha,\beta}(x - e_i) \tag{7}$$

where $\{e_i\}_{i=1}^{N}$ are error samples, $G_{\alpha,\beta}(\cdot)$ is a generalized Gaussian density (GGD) function:

$$G_{\alpha,\beta}(e) = \frac{\alpha}{2\beta \cdot \Gamma(1/\alpha)} \exp\left(-\left|\frac{e}{\beta}\right|^\alpha\right) \tag{8}$$

where $\alpha > 0$ and $\beta > 0$ refer to the kernel shape and bandwidth characteristics.

$\Gamma(a) = \int_0^\infty x^{a-1} e^{-x} dx, a > 0$ stands for Gamma function.

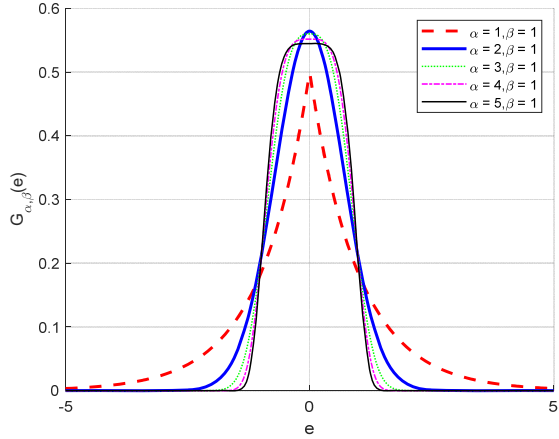

**Fig. 2** Generalized Gaussian density (GGD) function.

As shown in Fig. 2, the GGD function can exhibit greater adaptability to unknown non-Gaussian noise than the Gaussian function for the same kernel width $\beta$. Moreover, it has been evidenced that when the shape parameter $\alpha = 2$, the GGD distribution mathematically converges to its Gaussian counterpart [43].

In this study, the proposed mixture generalized Gaussian kernel through the mixture factor can be denoted as:

$$\begin{aligned} G_{\alpha,\beta}(\mathbf{e}) &= \eta G_{\alpha_1,\beta_1}(\mathbf{e}) + (1-\eta) G_{\alpha_2,\beta_2}(\mathbf{e}) \\ &= \eta \left( \frac{\alpha_1}{2\beta_1 \Gamma(1/\alpha_1)} \exp\left(-|\mathbf{e}/\beta_1|^{\alpha_1}\right) \right) \\ &\quad + (1-\eta) \left( \frac{\alpha_2}{2\beta_2 \Gamma(1/\alpha_2)} \exp\left(-|\mathbf{e}/\beta_2|^{\alpha_2}\right) \right) \end{aligned} \quad (9)$$

where $0 \leq \eta \leq 1$ is the mixture factor, which can adjust the proportion of two different kernels to maintain estimation accuracy. $\alpha_i, \beta_i$ $(i = 1, 2)$ are the shape parameter and bandwidth parameter of the $i^{th}$ generalized Gaussian kernel function.

Following Eq. (6), the estimation of the second-order information potential $\hat{V}_2(e)$ ($\alpha = 2$) in Eq. (5) can be obtained in the following expression:

$$\hat{V}_2(e) = \frac{1}{N} \sum_{i=1}^N p(e_i) = \frac{1}{N^2} \sum_{i=1}^N \sum_{j=1}^N G_{\alpha,\beta}(e_i - e_j) \quad (10)$$

Given that the function $(-\log)$ in Eq. (5) is monotonically decreasing, it is evident that the information potential $\hat{V}_2(e)$ should be maximized to obtain the minimum value of error entropy

$H_2(e)$.

## 3 SOC estimation via the TSGA with GMMEE-SRCKF

### 3.1 *Derivation of GMMEE-SRCKF*

This section will develop the GMMEE criterion with the SRCKF to obtain optimal accuracy in SOC estimation. The following part presents a detailed derivation of the GMMEE-SRCKF.

Step 1. Initialization:

$$\begin{cases} \hat{\mathbf{x}}_{0|0} = E[\mathbf{x}_0] \\ \xi_i = \sqrt{n}[\mathbf{1}]_i, \quad i = 1, 2, \ldots, 2n \end{cases} \tag{11}$$

where $E[\cdot]$ is the expectation operator, $[\mathbf{1}] \in \mathbf{R}_n$.

$$[\mathbf{1}] = \left\{ \begin{pmatrix} 1 \\ 0 \\ \vdots \\ 0 \end{pmatrix}, \begin{pmatrix} 0 \\ 1 \\ \vdots \\ 0 \end{pmatrix}, \cdots, \begin{pmatrix} 0 \\ 0 \\ \vdots \\ 1 \end{pmatrix}, \begin{pmatrix} -1 \\ 0 \\ \vdots \\ 0 \end{pmatrix}, \begin{pmatrix} 0 \\ -1 \\ \vdots \\ 0 \end{pmatrix}, \cdots, \begin{pmatrix} 0 \\ 0 \\ \vdots \\ -1 \end{pmatrix} \right\}$$

Step 2. Calculate the set of initial cubature points:

$$\begin{cases} \mathbf{P}_{\tau-1|\tau-1} = \mathbf{B}_{p,\tau-1}\mathbf{B}_{p,\tau-1}^T \\ \mathbf{X}_{i,\tau-1|\tau-1} = \mathbf{B}_{p,\tau-1}\xi_i + \hat{\mathbf{x}}_{\tau-1|\tau-1}, \quad i = 1, 2, \ldots, 2n \end{cases} \tag{12}$$

where $\mathbf{B}_{p,\tau-1}$ is the error covariance matrix obtained from Cholesky decomposition at time $\tau-1$, then the prior state $\hat{\mathbf{x}}_{\tau|\tau-1}$ is calculated by:

$$\mathbf{X}_{i,\tau|\tau-1}^* = f\left(\mathbf{X}_{i,\tau-1|\tau-1}, I_{\tau-1}\right) \tag{13}$$

$$\hat{\mathbf{x}}_{\tau|\tau-1} = \frac{1}{2n}\sum_{i=1}^{2n}\mathbf{X}_{i,\tau|\tau-1}^* \tag{14}$$

where $I_{\tau-1}$ is the current at time $\tau-1$, $f(\bullet)$ is the battery state Eq. (3).

Now, the covariance matrix $\mathbf{P}_{\tau|\tau-1}$ is given by:

$$\chi_{\tau|\tau-1}^* = \frac{1}{\sqrt{2n}}[\mathbf{X}_{1,\tau|\tau-1}^* - \hat{\mathbf{x}}_{\tau|\tau-1}, \cdots, \mathbf{X}_{2n,\tau|\tau-1}^* - \hat{\mathbf{x}}_{\tau|\tau-1}] \tag{15}$$

$$\mathbf{B}_{p,\tau|\tau-1} = \text{Tria}[\chi_{\tau|\tau-1}^*, \mathbf{B}_{Q,\tau-1}] \tag{16}$$

$$\mathbf{P}_{\tau|\tau-1} = \mathbf{B}_{p,\tau|\tau-1}\mathbf{B}_{p,\tau|\tau-1}^T, \quad \mathbf{Q}_{\tau-1} = \mathbf{B}_{Q,\tau-1}\mathbf{B}_{Q,\tau-1}^T \tag{17}$$

where Tria[•] denotes the general triangularization algorithm.

Step 3. Measurement update: After the time update, the measurement cubature points and predicted measurement can be computed by:

$$\mathbf{X}_{i,\tau|\tau-1} = \mathbf{B}_{p,\tau|\tau-1}\xi_i + \hat{\mathbf{x}}_{\tau|\tau-1} \tag{18}$$

$$\mathbf{Y}_{i,\tau|\tau-1} = h(\mathbf{X}_{i,\tau|\tau-1}, I_\tau) \tag{19}$$

$$\hat{\mathbf{y}}_{\tau|\tau-1} = \frac{1}{2n}\sum_{i=1}^{2n}\mathbf{Y}_{i,\tau|\tau-1} \tag{20}$$

where $I_\tau$ is the current at time $\tau$, $h(\bullet)$ is the observation Eq. (4).

Now, the covariance matrix $\mathbf{P}_{xy,\tau|\tau-1}$ can be expressed as:

$$\boldsymbol{\zeta}_{\tau|\tau-1} = \frac{1}{\sqrt{2n}}[\mathbf{Y}_{1,\tau|\tau-1} - \hat{\mathbf{y}}_{\tau|\tau-1}, \cdots, \mathbf{Y}_{2n,\tau|\tau-1} - \hat{\mathbf{y}}_{\tau|\tau-1}] \tag{21}$$

$$\mathbf{P}_{xy,\tau|\tau-1} = \boldsymbol{\chi}^*_{\tau|\tau-1}\boldsymbol{\zeta}^T_{\tau|\tau-1} \tag{22}$$

Step 4. Fixed-point iterative algorithm:

$$\overline{\mathbf{H}}_\tau = \mathbf{P}^T_{xy,\tau|\tau-1}\mathbf{P}^{-1}_{\tau|\tau-1} \tag{23}$$

$$\mathbf{D}_\tau = \mathbf{W}_\tau \mathbf{x}_\tau + \mathbf{e}_\tau \tag{24}$$

where

$$\mathbf{D}_\tau = \mathbf{B}^{-1}_\tau \begin{bmatrix} \hat{\mathbf{x}}_{\tau|\tau-1} \\ \mathbf{y}_\tau - \hat{\mathbf{y}}_{\tau|\tau-1} + \overline{\mathbf{H}}_\tau\hat{\mathbf{x}}_{\tau|\tau-1} \end{bmatrix} \tag{25}$$

$$\mathbf{W}_\tau = \mathbf{B}^{-1}_\tau \begin{bmatrix} \mathbf{I}_n \\ \overline{\mathbf{H}}_\tau \end{bmatrix} \tag{26}$$

$$\mathbf{e}_\tau = \mathbf{B}^{-1}_\tau \begin{bmatrix} -(\mathbf{x}_\tau - \hat{\mathbf{x}}_{\tau|\tau-1}) \\ \mathbf{r}(\tau) \end{bmatrix} \tag{27}$$

where $\mathbf{I}_n$ is $n \times n$ identity matrix, $\mathbf{y}_\tau$ denotes the actual measured terminal voltage of the circuit.

Meanwhile, $\mathbf{B}_\tau$ can be calculated by the Cholesky decomposition.

$$\begin{bmatrix} \mathbf{P}_{\tau|\tau-1} & 0 \\ 0 & \mathbf{R}_\tau \end{bmatrix} = \begin{bmatrix} \mathbf{B}_{p,\tau|\tau-1}\mathbf{B}^T_{p,\tau|\tau-1} & 0 \\ 0 & \mathbf{B}_{r,\tau}\mathbf{B}^T_{r,\tau} \end{bmatrix} = \mathbf{B}_\tau\mathbf{B}^T_\tau \tag{28}$$

According to Eq. (9) and Eq. (10), the cost function of the proposed method is defined as:

$$J_L(\mathbf{x}_\tau) = \frac{1}{L^2}\sum_{i=1}^{L}\sum_{j=1}^{L}\begin{bmatrix} \eta G_{\alpha_1,\beta_1}(\mathbf{e}_j - \mathbf{e}_i) \\ +(1-\eta)G_{\alpha_2,\beta_2}(\mathbf{e}_j - \mathbf{e}_i) \end{bmatrix} \tag{29}$$

where $L = n + m$.

The optimal state variable estimation must be achieved by maximizing the objective function to attain the maximum value.

$$\begin{aligned}\hat{\mathbf{x}}_\tau &= \arg\max_{\mathbf{x}_\tau} J_L(\mathbf{x}_\tau) \\ &= \arg\max_{\mathbf{x}_\tau} \frac{1}{L^2}\sum_{i=1}^{L}\sum_{j=1}^{L}\begin{bmatrix} \eta G_{\alpha_1,\beta_1}(\mathbf{e}_j - \mathbf{e}_i) \\ +(1-\eta)G_{\alpha_2,\beta_2}(\mathbf{e}_j - \mathbf{e}_i) \end{bmatrix}\end{aligned} \tag{30}$$

The posterior state estimation can be obtained by setting the gradient of the cost function $J_L(\mathbf{x}_\tau)$ with respect to $\mathbf{x}_\tau$ to zero. Since these steps are lengthy but straightforward, they are omitted here for brevity. The complete mathematical derivation is provided in the Appendix for reference. Finally, the posterior state estimation and covariance matrix are presented by:

$$\hat{\mathbf{x}}_\tau = \hat{\mathbf{x}}_{\tau|\tau-1} + \hat{\mathbf{K}}_\tau \left( y_\tau - \bar{\mathbf{H}}_\tau \hat{\mathbf{x}}_{\tau|\tau-1} \right) \tag{31}$$

$$\mathbf{P}_{\tau|\tau} = \left( \mathbf{I_n} - \hat{\mathbf{K}}_\tau \bar{\mathbf{H}}_\tau \right) \mathbf{P}_{\tau|\tau-1} \left( \mathbf{I_n} - \hat{\mathbf{K}}_\tau \bar{\mathbf{H}}_\tau \right)^T + \hat{\mathbf{K}}_\tau \mathbf{R}_\tau \left( \hat{\mathbf{K}}_\tau \right)^T \tag{32}$$

At this point, the proposed GMMEE-SRCKF has been derived. Under conditions involving outlier-induced non-Gaussian noise complexities, MEE-CKF fails to achieve optimal estimation. To address these limitations, an integration of GMMEE with SRCKF has been proposed. It can be demonstrated that by leveraging two adjustable error entropies, the proposed mixture factor-enhanced cost function achieves robust SOC estimation against measurement outliers.

*Remark* 1. It can be deduced that the GMMEE criterion will degenerate into the GMEE criterion: identity of kernel bandwidth configurations ($\beta_1 = \beta_2$), or if the mixing factor $\eta = 0$ or 1. Additionally, the generalized minimum error entropy criterion-based cubature Kalman filter (GMEE-CKF) can still outperform the MEE-CKF with proper parameters and kernel width $\beta_1$ or $\beta_2$.

*Remark* 2. When $\alpha_1 = \alpha_2 = 2$, the GMMEE criterion is reduced to the MMEE criterion. As one of the specified forms of GMMEE, MMEE can develop robust performance with proper mixture kernel functions $G_{\beta_1}(\bullet)$ and $G_{\beta_2}(\bullet)$. This is because the cost function constructed by two error entropies can be adapted to a wider range of practical SOC estimation, enhancing the efficacy of the MMEE criterion-based CKF.

*Remark* 3. Furthermore, the MMEE criterion can be seen as a mixture of the MEE. As mixture factor $\eta = 0$ or 1, the MMEE criterion-based CKF will reduce to the MEE-CKF with one kernel function $G_{\beta_1}(\bullet)$ or $G_{\beta_2}(\bullet)$.

**3.2 *Kernel parameter optimization via TSGA hybrid optimization***

To further improve the performance of the proposed GMMEE-SRCKF, a hybrid optimization algorithm based on the tree seed algorithm (TSA) [46,47] and genetic algorithm (GA) (TSGA) is introduced to optimize the kernel parameters, including the shape parameters and the bandwidth parameters $\alpha_i, \beta_i$ ($i = 1, 2$).

The hybrid TSGA algorithm combines the efficient local convergence characteristics of TSA with the robust global search capabilities of GA, enabling fast and reliable convergence to the global optimum. This collaborative mechanism facilitates more effective optimization of the GMMEE kernel parameters, achieving an improved balance between convergence speed and global search performance.

Two update rules are employed to generate new seeds:

$$Z_{i,j} = w * R_{i,j} + \sigma_{i,j} * (B_j - R_{r,j}) \tag{33}$$

$$Z_{i,j} = w * R_{i,j} + \sigma_{i,j} * (R_{i,j} - R_{r,j}) \tag{34}$$

where $Z_{i,j}$ denotes the $j$-th dimension of the new seed generated by the individual $i$. $R_{i,j}$ represents the current value of the $j$-th dimension of the individual $i$. $R_{r,j}$ is the value of the $j$-th dimension of a randomly selected peer individual $r$. $B_j$ indicates the $j$-th dimension of the best individual found so far. $w$ is the inertia weight, controlling the balance between the current position and the new seed. A larger $w \in (0.7, 1)$ favors broader exploration by retaining more of the current position, while a smaller $w \in (0.3, 0.6)$ emphasizes local exploitation by allowing finer adjustments in the later search stages, thereby enhancing convergence accuracy. Typically, $w$ is set larger in the early and decreased in the later iterations. $-1 \leq \sigma_{i,j} \leq 1$ is a randomly generated learning coefficient for individual $i$ at dimension $j$. This factor enhances search diversity by enabling movements towards or away from reference solutions. Eq. (33) emphasizes learning from the best individual found so far, while Eq. (34) promotes exploration by considering the difference between the current individual and a randomly selected peer.

The generation process of the seeds is critical for the identification of optimal values. However, the generation of seeds based on Eq. (34) is extremely simple, which may have a significant impact on the results. In contrast, the crossover and mutation operations of the GA explicitly introduce new genetic combinations and random alterations at the population level. By replacing the global random seed generation in the TSA with genetic algorithm-based crossover and mutation operators, the hybrid approach significantly enhances population diversity and exploration capability, thereby avoiding premature convergence and improving robustness. Furthermore, a search parameter $ST \in [0, 1]$ is utilized to regulate exploration scope. Smaller values may introduce excessive perturbations due to GA, whereas larger values tend to increase the risk of premature convergence due to TSA. It is demonstrated by experimental results that the effective range of ST is set to [0.5,0.7].

Meanwhile, the initial population of trees is generated within the search space:

$$R_{i,j} = L_{j,\min} + r_{i,j}(H_{j,\max} - L_{j,\min}) \tag{35}$$

where $R_{i,j}$ denotes the positional coordinates of the $i$-th tree in the $j$-th dimension, $L_{j,\min}$ and $H_{j,\max}$ represent the lower and upper bounds of the search space, respectively, and $r_{i,j}$ is a Uniformly distributed random variable within [0, 1]. For minimization problems, Eq. (36) identifies the optimal tree position:

$$B = \arg\min\{f(\tilde{R}_i)\}, \quad i = 1, 2, \ldots, N \tag{36}$$

where N is the initial population of the trees, and $f(\tilde{R}_i)$ evaluates the fitness value of the $i$-th tree, while the fitness value denotes the root mean square error (RMSE) of SOC estimation under specified

---

**Algorithm 1:** TSGA hybrid optimization

**Input:** the shape parameters and the bandwidth parameters ($\alpha_1, \alpha_2, \beta_1, \beta_2$)
    Population size (N)
    Maximum number of iterations (Max_iter)
    Inertia weight ($w$)
    Search parameter (ST)
**Output:** Optimized kernel parameters for GMMEE-SRCKF

```
1:  begin
2:      Generate an initial population of N individuals using Eq. (35)
3:      Evaluate the fitness (RMSE of SOC estimation) of each individual
4:      Select the initial global best solution using Eq. (36)
5:      Set iteration counter t = 0
6:      while (t < Max_iter)
7:          for each individual (tree)
8:              Generate a number of seeds for this tree (10% to 25% of the population size)
9:              for each seed
10:                 for each dimension
11:                     if (rand < ST)
12:                         Update this dimension using Eq. (33)
13:                     else
14:                         Update this dimension by crossover and mutation operation
15:                     end if
16:                     Constrain the updated dimension within the search space
17:                 end for
18:             end for
19:             Evaluate all seeds and select the best seed
20:             if (best seed is better than current tree)
21:                 Replace the tree with the best seed
22:             end if
23:         end for
24:         if (the current best solution is better than the stored best solution) then
25:             Update the global best solution
26:         end if
27:         Increment t = t + 1
28:     end while
29:     Return the global best solution
30: end
```

experimental conditions. The pseudo-code of the TSGA method is summarized as Algorithm 1.

### 3.3 *SOC estimation based on TSGA-optimized GMMEE-SRCKF*

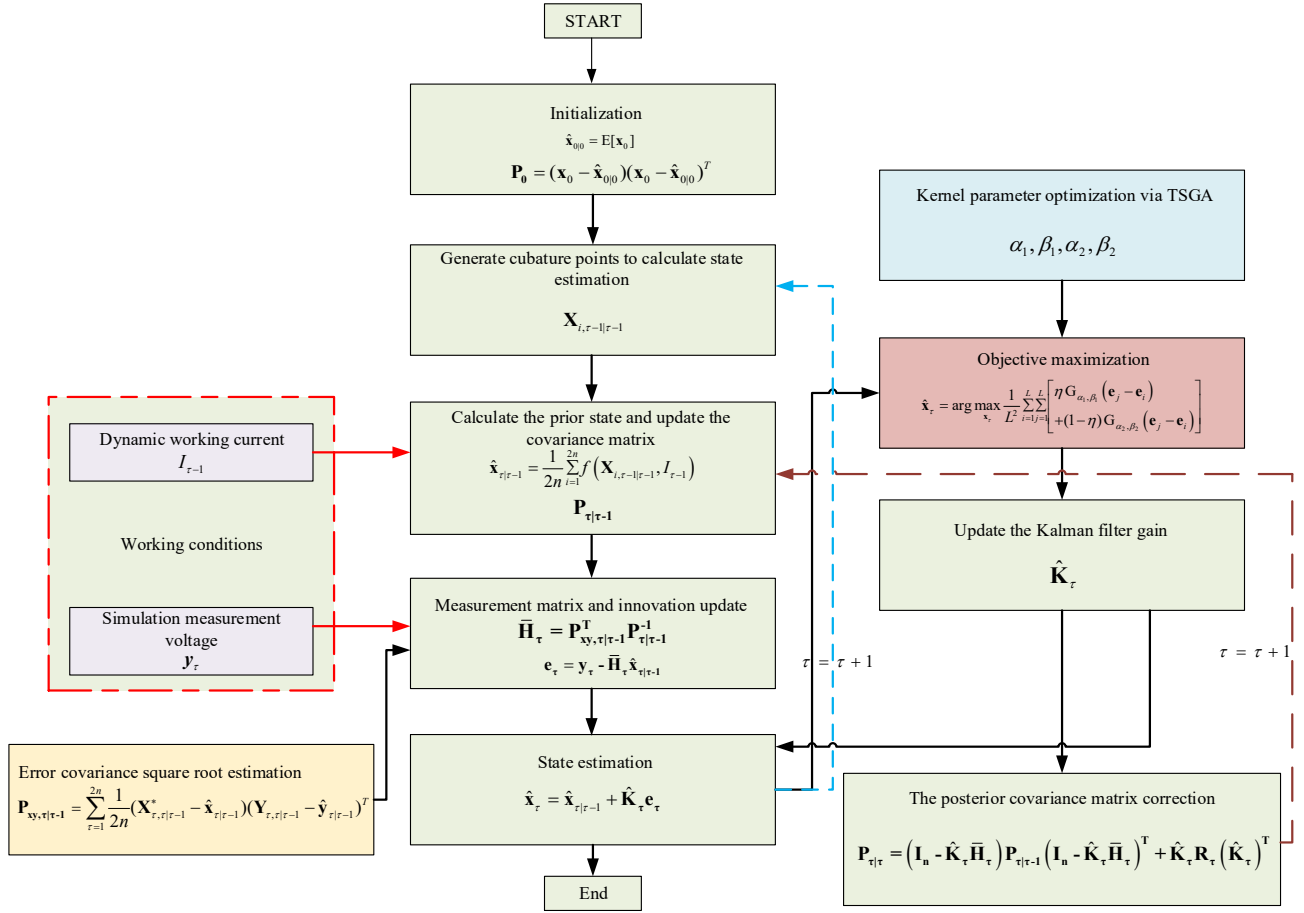

**Fig. 3** Flowchart of TSGA-optimized GMMEE-SRCKF.

The flowchart of the TSGA-optimized GMMEE-SRCKF is presented in Fig. 3. While the SOC estimation steps are specifically outlined below:

Step 1. Battery dataset. To verify the robustness of the algorithm, experiments should be performed in multiple environments. Concurrently, the requisite data, including voltage, current, and temperature, can be obtained from the relevant experiments. Additionally, the data will be obtained from the publicly accessible battery dataset.

Step 2. Identification of corresponding parameters. Offline identification is used to calculate the battery parameters. It is imperative to discharge the battery under varying conditions and establish the SOC-OCV relationship. Subsequently, the battery data must be integrated with the SOC-OCV relationship to calculate corresponding parameters.

Step 3. Filter design. To evaluate the efficacy and precision of TSGA-optimized GMMEE-

SRCKF for estimating the SOC under complex non-Gaussian noise. Accordingly, the MD-based method will be employed to verify the robustness of the TSGA-optimized GMMEE-SRCKF.

Step 4. Experiments in varying conditions. It is imperative to investigate the robustness of the TSGA-optimized GMMEE-SRCKF in various complex working conditions. Therefore, simulation experiments need to be carried out under a variety of operating conditions.

Step 5. Experimental results analysis. Given the series of specific experiments that have been conducted, the robustness of the TSGA-optimized GMMEE-SRCKF is evidenced through the obtained experimental outcomes.

# 4 Experiments and discussion

## 4.1 *Battery dataset*

This paper employed two different battery datasets. The first was obtained from a new SAMSUNG INR21700-30T lithium-ion cell with a nominal capacity of 3 Ah [53]. Testing was carried out in an 8-cubic-foot thermal chamber using a Digatron firing circuits universal battery tester, whose channels support currents up to 75 A and voltages up to 5 V, with a voltage sampling frequency of 10 Hz and a voltage and current measurement accuracy of 0.1% of full scale. The second dataset was a publicly available ageing dataset reported in scientific data [54], comprising twenty commercial 18650 LiFePO$_4$ cells that had experienced a 40% capacity loss. All cells were welded to printed circuit boards with four-point measurement capability, with two terminals dedicated to current injection and two to voltage sensing, to minimise contact resistance effects. The aged battery dataset utilised in this work had undergone 3100 cycles at a controlled temperature of 50 °C, maintained using laboratory ovens, resulting in a substantial capacity reduction from the initial 1.1 Ah to 0.65 Ah. During the cycle testing, cell measurement voltage was recorded at a sampling frequency of 1 Hz. More details can be found in Table 1.

**Table 1**
Details of LiNCM and LiFePO$_4$ batteries.

| Type | Material | Nominal Capacity (Ah) | Nominal voltage (V) | Max operating voltage (V) | Min operating voltage (V) |
|---|---|---|---|---|---|
| INR21700 | LiNMC | 3 | 3.6 | 4.2 | 2.5 |
| 18650 | LiFePO$_4$ | 1.1 | 3.2 | 3.65 | 2 |

In this study, the INR21700 dataset was obtained from Lithium-ion batteries operated at three temperatures under two operational conditions. The first working condition is the Urban Dynamometer Driving Schedule (UDDS) cycle, which represents urban driving conditions. The second working condition is the Urban Supplemental Federal Test Procedure 06 (US06) cycle. The US06 Driving Cycle has been designed to simulate the characteristics of short time, high speed, and sudden acceleration or deceleration. This is intended to more accurately reflect the demanding driving conditions encountered on highways. The aforementioned two working conditions will be conducted at temperatures of 40°C, 25°C, and -10°C. The current, voltage, and SOC values will be presented in Fig. 4a and 4b. All the parameters and SOC curves used in this paper were taken from [53], which has been validated by [26]. It is noteworthy that the ageing dataset was evaluated using the Worldwide Harmonized Light Vehicles Test Procedure (WLTP) cycle, which simulates a 30-minute current discharge of real-world electric vehicle operation.

Consequently, the SOC was calculated using the ampere-hour integration method, as expressed in Eq. (3). To account for battery lifespan considerations, the SOC of battery packs in electric vehicles is typically constrained above a minimum threshold. In this study, the threshold is set to 10% for the INR21700 dataset, and data below this level were excluded from SOC estimation. Meanwhile, the WLTP cycle test was carried out with a depth of discharge limited to 70%, so the battery is discharged only down to 30% SOC during the test, ensuring both safe operation and realistic evaluation conditions.

As for the INR21700 dataset, the SOC–OCV relationship was extracted from HPPC tests performed at discrete SOC levels between 100% and 10%, and a 6th-order polynomial regression model was adopted for characterizing the nonlinear SOC-OCV mapping. Meanwhile, the fitting curves at the three specified temperatures are shown in Fig. 4c. In addition, the SOC–OCV curve associated with the WLTP cycle was derived from characterisation measurements carried out at 50 °C.

Before conducting the SOC estimation experiments, the second-order RC ECM was identified using the acquired dataset. The accuracy of the model was validated by comparing the simulated terminal voltage with the measured voltage. As demonstrated by the INR21700 dataset, the voltage identification results at three temperatures are presented in Fig. 5 and the simulated voltage closely mirrors the measured voltage. It is noteworthy that under low-temperature conditions, the polarization effect becomes significantly intensified, resulting in pronounced variations in parameters. In Table 2, the maximum absolute error (MAX), mean absolute error (MAE), and mean square error (MSE) are employed to evaluate the accuracy of parameter identification. The maximum error occurs under the extreme temperature condition, where the RMSE reaches 0.0398 V and the MAE is 0.0281 V. This increase in error is primarily attributed to the pronounced temperature dependence of battery's internal resistance and polarization effects. Nevertheless, the accuracy remains within an acceptable range for subsequent SOC estimation.

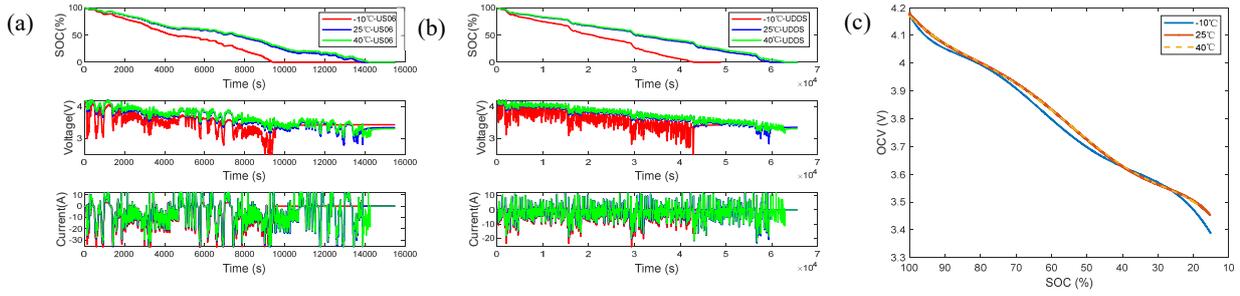

**Fig.** 4 (a) Dataset obtained from US06 Driving Cycle (b) dataset obtained from UDDS working conditions (c) curve of SOC-OCV.

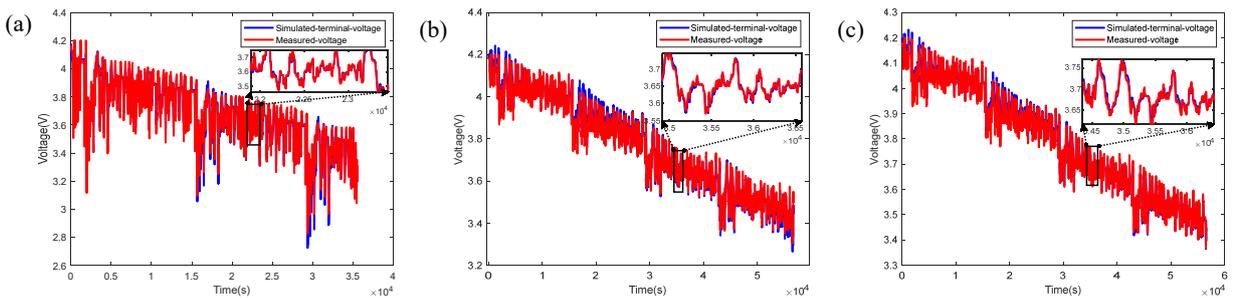

**Fig. 5** Terminal voltage comparison across three temperatures under the UDDS working condition: (a) -10℃ (b) 25℃ (c) 40℃.

**Table 2**
Error of simulated voltage under varying working condition.

| Error/Working Condition | UDDS -10°C | UDDS 25°C | UDDS 40°C | WLTP 50°C |
|---|---|---|---|---|
| MAX(V) | 0.2829 | 0.0604 | 0.0789 | 0.0539 |
| MAE(V) | 0.0253 | 0.0155 | 0.0123 | 0.0281 |
| RMSE(V) | 0.0398 | 0.0208 | 0.0165 | 0.0361 |

The experiments will be conducted under various conditions, including complex non-Gaussian noise, varying temperatures, diverse working conditions, aged batteries and different initial values. Additionally, we will analyze the elapsed time of the proposed algorithm and compare its performance with other prominent algorithms. All experiments were carried out on an Intel i7-14650HX 2.20 GHz running MATLAB 2023a. Notably, all GMMEE kernels parameters in the proposed algorithm were optimized through the TSGA method. In addition to the RMSE, the MSE and MAE will also be utilized to quantitatively evaluate the estimation accuracy of the SOC.

### 4.2 Different non-Gaussian noises

The experiments will be carried out under Laplace mixed noise and Uniformly mixed noise, and both experiments will be conducted under the UDDS condition at 25°C. Both mixed noises are constructed as $\rho(t) = (1-\gamma(t))\alpha(t) + \gamma(t)\beta(t)$, $\gamma(t)$ denotes the weight of the each noise, while $P\{\gamma(t)=0\}=1-c$ represents the probability of noise $\alpha(k)$, $P\{\gamma(t)=1\}=c$ represents the probability of noise $\beta(t)$, where $c$ is generated by a binary noise function. In this article, $c$ is set as 0.95. In the two types of mixed noise, $\alpha(t)$ denotes Gaussian noise, for which the mean value is 0.1 and the variance is 10, while $\beta(t)$ denotes two complex non-Gaussian noises: Laplace noise, for which the mean value is 1 and the variance is 1, and the Uniform distribution of noise with an interval of $[-4, 2]$. The two noises are illustrated in Fig. 6a and 6b. In this section, two working conditions have the same process noise covariance with $\mathbf{Q} = 1 \times 10^{-6} \times \mathbf{I}_3$, where $\mathbf{I}_3$ is $3 \times 3$ identity matrix, with prior error covariance $\mathbf{P}_{0|0} = diag[0.01, 0.01, 0.06]$.

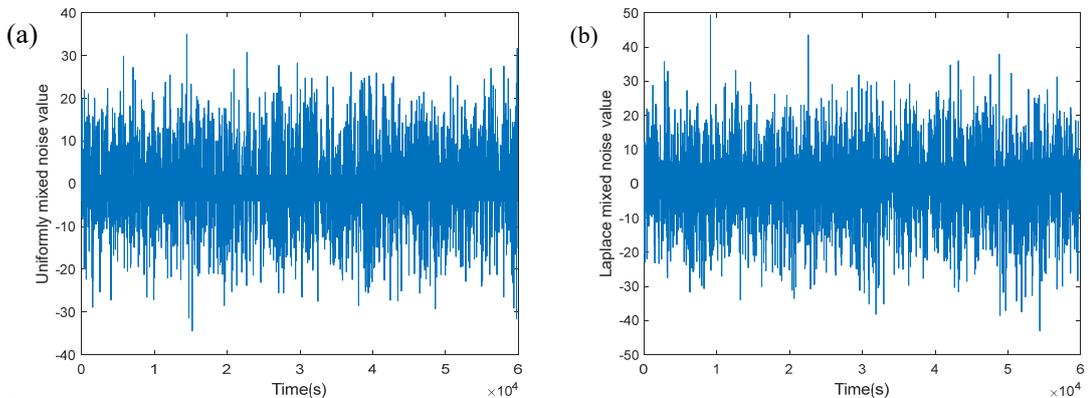

**Fig. 6** (a) Uniformly mixed measurement noise (b) Laplace mixed measurement noise.

### 4.2.1 *Uniformly mixed noise*

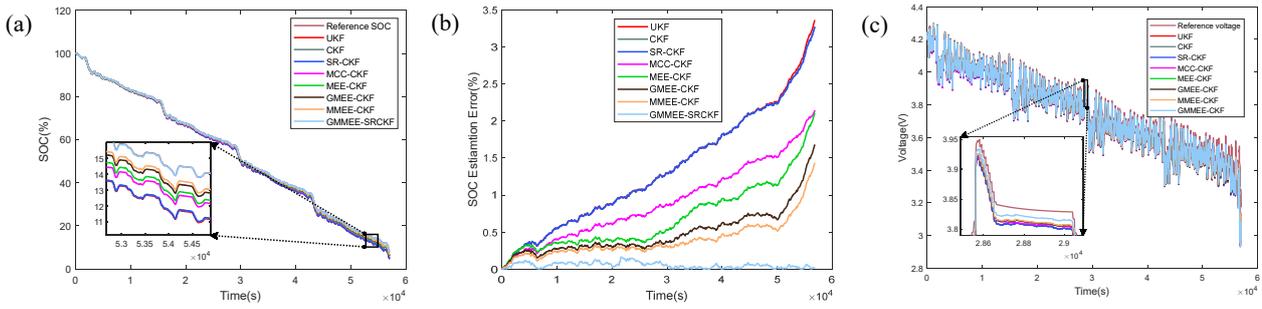

**Fig. 7** (a) SOC estimation results under Uniformly mixed noise (b) absolute error (c) voltage estimation result.

In this subsection, the experiments are carried out under Uniformly mixed noise, while the optimized kernel parameters are set as $\eta = 0.5$, $\alpha_1 = 3.3662$, $\alpha_2 = 4.3453$, $\beta_1 = 0.0007788$, $\beta_2 = 0.0000983$. As shown in Fig. 7a and 7b, the various algorithms can be observed and compared with each other, while they depict the SOC estimation results and error. In addition, Fig. 7c presents the estimated voltages obtained by different algorithms along with the reference terminal voltage, showing that the proposed algorithm can track the reference voltage well. The accuracy of the predicted voltage is crucial, as it directly affects the innovation term and thereby influences the posterior SOC update in the Kalman filter. As we all know, the CKF demonstrates superior performance compared to the UKF in general, and the results presented in this figure provide further evidence to support it. However, it was found that the original MMSE-based filter exhibited reduced robustness under non-Gaussian noise and tended to generate larger estimation errors. Compared to the original filters, robust algorithm-based KFs all exhibit greater accuracy than the above two methods. By contrast, the proposed GMMEE-SRCKF estimation method presents the lowest level of error under Uniformly mixed noise due to the usage of the GMMEE, which has more flexible kernel width options. To obtain a more detailed comparison of the numerical errors, the evaluation indices are listed in Table 3.

**Table 3**
Results of MAE, MSE, and RMSE at Uniformly mixed noise.

|  | Uniformly mixed noise | | |
| --- | --- | --- | --- |
|  | MAE | MSE | RMSE |
| UKF | 1.3428 | 2.4371 | 1.5611 |
| CKF | 1.3348 | 2.3961 | 1.5479 |
| SRCKF | 1.3340 | 2.3929 | 1.5469 |
| MCC-CKF | 0.8480 | 0.9491 | 0.9742 |
| MEE-CKF | 0.6980 | 0.6869 | 0.8288 |
| GMEE-CKF | 0.4895 | 0.3331 | 0.5772 |
| MMEE-CKF | 0.3933 | 0.2152 | 0.4639 |
| **GMMEE-SRCKF** | **0.0533** | **0.0042** | **0.0645** |

### 4.2.2 *Laplace mixed noise*

In this subsection, the experiments will be conducted to compare the robustness of the GMMEE-SRCKF with the other methods under Laplace mixed noise. In this experiment, the optimized kernel parameters are set as $\eta = 0.5$, $\alpha_1 = 2.324$, $\alpha_2 = 2.1413$, $\beta_1 = 0.0004122$, $\beta_2 = 0.00005$. The experimental results are illustrated in Figs. 8a-8c. The traditional CKF and UKF are still worse than the other methods due to the existence of impulse noise, while the proposed GMMEE-SRCKF method achieves the minimum amount of error, which means that this method could maintain good traceability and excellent estimation accuracy. As demonstrated in Table 4, the error analysis indicates that the GMMEE-SRCKF constitutes an optimal method for addressing complex non-Gaussian processes in SOC estimation.

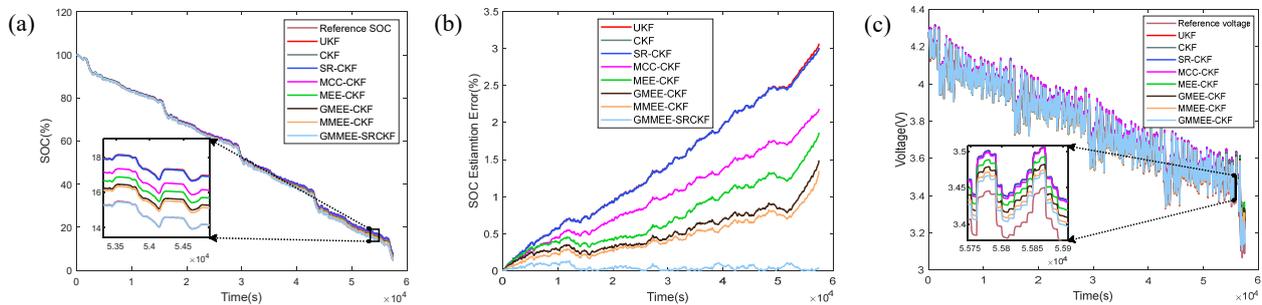

**Fig. 8** (a) SOC estimation results under Laplace mixed noise (b) absolute error (c) voltage estimation result.

**Table 4**
Results of MAE, MSE, and RMSE at Laplace mixed noise.

|  | Laplace mixed noise | | |
| --- | --- | --- | --- |
|  | MAE | MSE | RMSE |
| UKF | 1.3418 | 2.4124 | 1.5532 |
| CKF | 1.3373 | 2.3926 | 1.5468 |
| SRCKF | 1.3365 | 2.3898 | 1.5459 |
| MCC-CKF | 0.9565 | 1.2113 | 1.1006 |
| MEE-CKF | 0.7154 | 0.6906 | 0.8310 |
| GMEE-CKF | 0.5149 | 0.3623 | 0.6019 |
| MMEE-CKF | 0.4130 | 0.2391 | 0.4890 |
| **GMMEE-SRCKF** | **0.0411** | **0.0027** | **0.0517** |

### 4.3 Computational time analysis

To provide a more detailed evaluation of the practical applicability of the GMMEE-SRCKF, the average and maximum elapsed times used by each algorithm during a single discharge cycle are calculated. The elapsed time of the serval algorithm is presented in Table 5. As demonstrated in the previous subsection, the accuracy of SOC estimation for different matrix decomposition methods is approximately equivalent. However, there is a significant difference in the speed of convergence. The square root converges faster than the Cholesky decomposition. Meanwhile, one could also see that robust algorithm-based KFs take more time than the original due to the fixed-point iteration algorithm. However, it should be noted that the sampling period of the battery is 0.1 s, and the elapsed time taken by the GMMEE-SRCKF is far less than the sampling time spent during a single discharge cycle. Moreover, the proposed GMMEE-SRCKF can still achieve excellent estimation accuracy while having almost the same elapsed time as the other methods. Therefore, we can conclude that the method we proposed is feasible for SOC estimation.

**Table 5**
Comparison of elapsed time under different noise.

|  | Uniformly mixed noise | | Laplace mixed noise | |
| --- | --- | --- | --- | --- |
|  | MAX (ms) | MEAN (ms) | MAX (ms) | MEAN (ms) |
| UKF | 6.2105 | 0.0302 | 8.5324 | 0.0333 |
| CKF | 4.7780 | 0.0311 | 6.8002 | 0.0341 |
| SRCKF | 2.8892 | 0.0255 | 2.1820 | 0.0285 |
| MCC-CKF | 7.7351 | 0.0705 | 9.6828 | 0.0784 |
| MEE-CKF | 8.6313 | 0.0841 | 8.9107 | 0.0910 |
| GMEE-CKF | 7.5040 | 0.0806 | 7.0209 | 0.0882 |
| MMEE-CKF | 6.8939 | 0.0803 | 7.7497 | 0.0889 |
| **GMMEE-SRCKF** | **6.9861** | **0.0929** | **6.9823** | **0.1033** |
| Time(ms) | 51.6281 | 0.4952 | 57.8610 | 0.5457 |

### 4.4 SOC estimation at different ambient temperatures

The experiment we have conducted above demonstrated the efficacy of the GMMEE-SRCKF in accurately tracking the reference SOC under normal temperatures. However, it is necessary to perform experiments at varying temperatures to verify the feasibility of the GMMEE-SRCKF in a more challenging environment. The experiments were conducted under two temperatures to further investigate the viability in practice. Both experiments are carried out under Laplace mixed noise and

UDDS working conditions. In this section, the optimized kernel parameters are set as $\eta = 0.5$, $\alpha_1 = 3.423$, $\alpha_2 = 3.121$, $\beta_1 = 0.00005216$, $\beta_2 = 0.000412$ under low temperature and $\eta = 0.5$, $\alpha_1 = 3.531$, $\alpha_2 = 2.9$, $\beta_1 = 0.00007321$, $\beta_2 = 0.0002352$ under high temperature.

#### 4.4.1 *Performance analysis under low temperature (-10°C)*

The evaluation values are presented in Table 6. According to the experimental results, it is obvious that the estimated value based on the proposed method can smoothly track the reference SOC compared to the others, even in harsh environments. The estimation results and absolute error are presented in Figs. 9a-9c, highlighting that the GMMEE-SRCKF achieves a notable improvement in accuracy over other algorithms. All the performance evaluations of the proposed method are less than 0.4% at -10°C. By contrast, the original KFs deteriorate rapidly. Due to the internal chemical reaction

**Table 6**
Results of MAE, MSE, and RMSE at -10°C under UDDS working condition.

|  | MAE | MSE | RMSE |
| --- | --- | --- | --- |
| UKF | 1.0088 | 1.9263 | 1.3879 |
| CKF | 1.0031 | 1.8902 | 1.3749 |
| SRCKF | 1.0029 | 1.8895 | 1.3746 |
| MCC-CKF | 0.8711 | 1.4378 | 1.1991 |
| MEE-CKF | 0.5965 | 0.6970 | 0.8349 |
| GMEE-CKF | 0.4109 | 0.3365 | 0.5801 |
| MMEE-CKF | 0.3342 | 0.1962 | 0.4429 |
| **GMMEE-SRCKF** | **0.3234** | **0.1394** | **0.3733** |

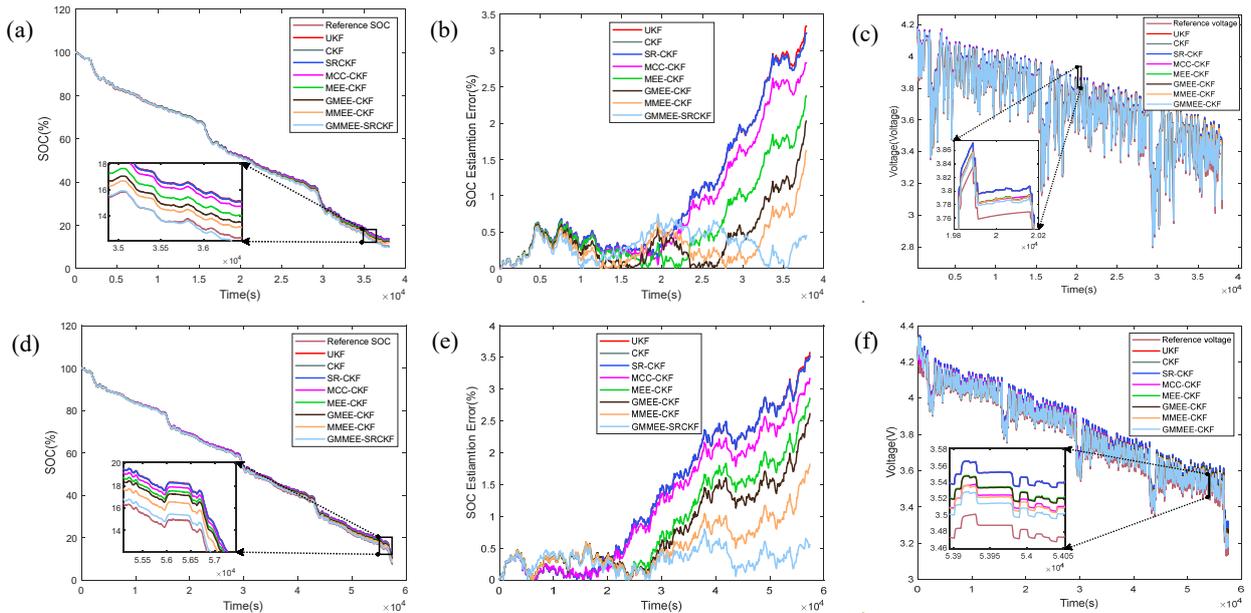

**Fig. 9** (a) SOC estimation result under low temperature (b) absolute error (c) voltage estimation result (d) SOC estimation results under high temperature (e) absolute error (f) voltage estimation result.

slowing down with decreasing temperature, Lithium-ion batteries perform poorly at low temperatures. Nevertheless, the proposed algorithm can still maintain optimal performance in this case, with maximum error evaluation metrics maintained at approximately 0.4%, which is within the acceptable range for SOC estimation. Therefore, this evidence provides compelling support for the robustness of the GMMEE-SRCKF.

**4.4.2 *Performance analysis under high temperature (40°C)***

As shown in Figs. 9e-9f, the results indicate that despite the battery working at high temperatures, the proposed algorithm can still optimize SOC estimation performance. The detailed performance is exhibited in Table 7. The error evaluation metrics of GMMEE-SRCKF at 40°C are still more numerically accurate than the other algorithms. The results presented in this subsection provide further evidence that the proposed method is capable of achieving good convergence and robustness in this case. Furthermore, the experimental results across different temperatures reveal that larger errors occur at lower temperatures rather than higher ones. Despite this, the GMMEE-SRCKF effectively enhances SOC estimation accuracy over a wide temperature range.

Table 7
Results of MAE, MSE, and RMSE at 40°C under UDDS working condition.

|  | MAE | MSE | RMSE |
| --- | --- | --- | --- |
| UKF | 1.3192 | 2.9412 | 1.7150 |
| CKF | 1.3164 | 2.9223 | 1.7095 |
| SRCKF | 1.3176 | 2.9280 | 1.7111 |
| MCC-CKF | 1.1791 | 2.3222 | 1.5239 |
| MEE-CKF | 0.9151 | 1.4061 | 1.1858 |
| GMEE-CKF | 0.7916 | 1.0314 | 1.0156 |
| MMEE-CKF | 0.5480 | 0.4464 | 0.6681 |
| **GMMEE-SRCKF** | **0.3171** | **0.1226** | **0.3501** |

**4.5 *Different initial values of SOC***

In this subsection, three initial SOC values (90%, 80%, and 70%) were tested under Uniformly mixed noise. As shown in Figs. 10a and 10b, when the initial SOC value deviates from the true value, the estimation process requires more time to converge. Additionally, SOC estimation under different initial values demonstrates an initial rising phase to maintain accuracy, indicating that initial SOC errors significantly impact the estimation process. Nevertheless, experimental findings demonstrate that the GMMEE-SRCKF exhibits excellent convergence performance, primarily attributed to its augmented efficacy in managing complex non-Gaussian noise. Moreover, the estimation error tends to increase proportionally with the initial SOC error, highlighting the importance of providing a more accurate initial value.

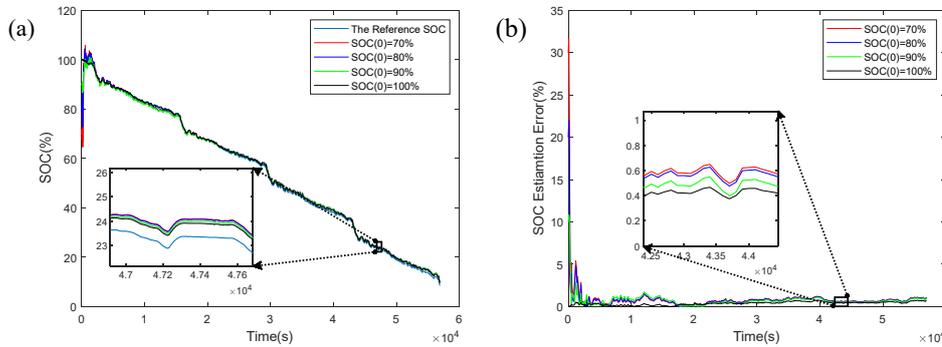

**Fig. 10** (a) SOC estimation with different initial values (b) absolute error with different initial values.

### 4.6 *SOC estimation at three temperatures under US06 testing*

The findings from the previous experiments confirm that the GMMEE-SRCKF effectively achieves accurate SOC estimation under UDDS conditions. However, Lithium battery operating conditions can be more intricate. This section evaluates the performance of GMMEE-SRCKF under US06 testing at -10°C, 25°C, and 40°C. In this section, the manually selected kernel parameters are set to $\eta = 0.5$, $\alpha_1 = 2.231$, $\alpha_2 = 1.9$, $\beta_1 = 0.00009143$, $\beta_2 = 0.00003$. The corresponding results, shown in Figs. 11a-11d, indicate that the proposed method delivers highly accurate SOC estimation.

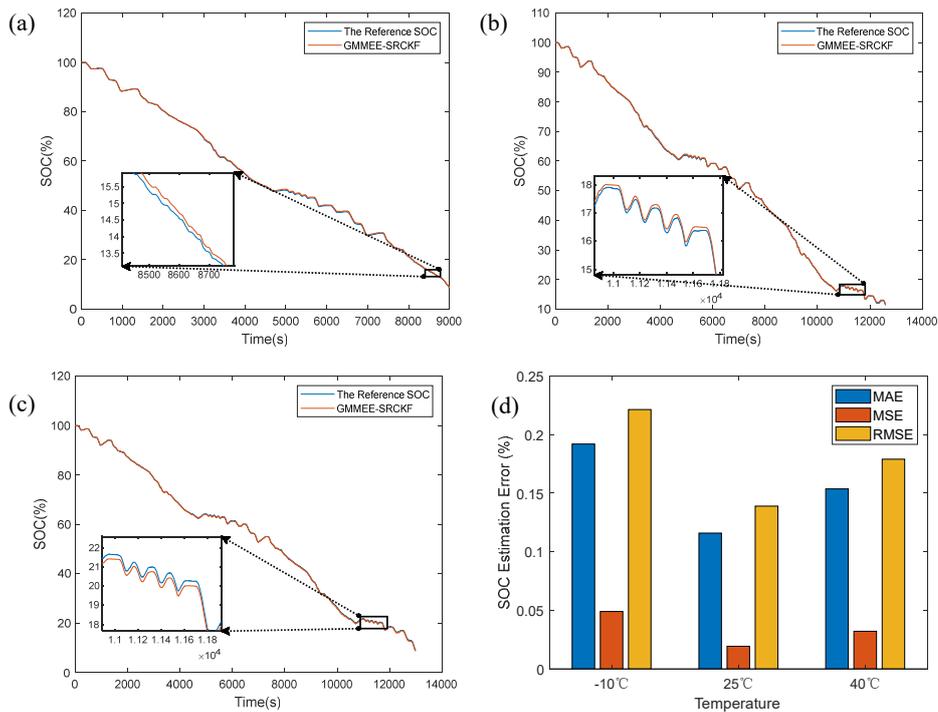

**Fig. 11** SOC estimation results under US06 working conditions: (a) -10 ℃ (b) 25°C (c) 40 °C (d) Results of evaluation indexing.

As reflected in the error curve, estimation errors tend to increase when temperatures deviate from room temperature, with a more pronounced rise at lower temperatures, aligning with the analysis from the previous section. The experimental results presented here further validate the capability of GMMEE-SRCKF to operate effectively under complex conditions and across a broad temperature range.

### 4.7 *SOC estimation under aging conditions*

Battery ageing leads to capacity degradation and an increase in internal resistance, both of which can affect SOC estimation accuracy. Therefore, this section further evaluates the performance of the proposed GMMEE-SRCKF algorithm under battery ageing conditions, in the presence of Uniformly mixed noise during the 50 °C WLTP cycle. In this section, the manually selected kernel parameters are set to $\eta = 0.5$, $\alpha_1 = 1.52332$, $\alpha_2 = 1.8707$, $\beta_1 = 0.00004143$, $\beta_2 = 0.00041299$. As shown in Table 8 and Fig. 12, GMMEE-SRCKF consistently exhibits superior SOC estimation accuracy compared to other algorithms, even under ageing conditions. Despite extensive cycling and exposure to extreme high temperatures, the error indexes remain significantly lower, highlighting the strong robustness and adaptability of the method to battery degradation.

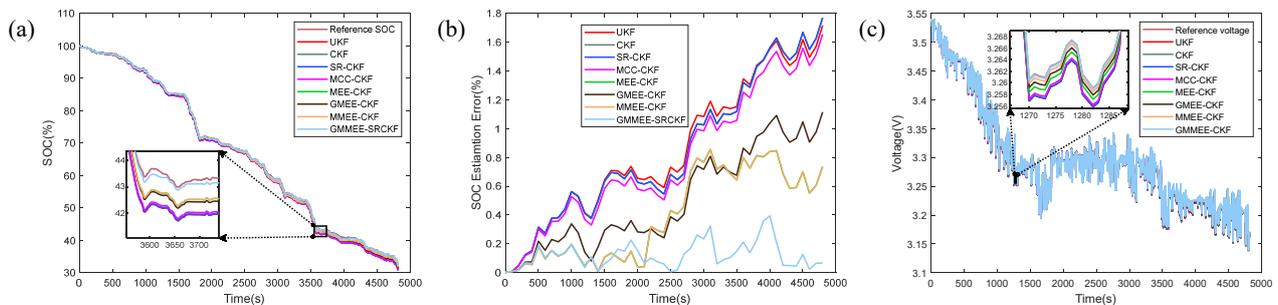

**Fig. 12** (a) SOC estimation result under aging conditions (b) absolute error (c) voltage estimation.

**Table 8**
Results of MAE, MSE, and RMSE under aging conditions.

|  | MAE | MSE | RMSE |
|---|---|---|---|
| UKF | 0.8587 | 0.9660 | 0.9828 |
| CKF | 0.8467 | 0.9548 | 0.9771 |
| SRCKF | 0.8470 | 0.9554 | 0.9774 |
| MCC-CKF | 0.7948 | 0.8431 | 0.9182 |
| MEE-CKF | 0.6619 | 0.5689 | 0.7543 |
| GMEE-CKF | 0.5079 | 0.3723 | 0.6101 |
| MMEE-CKF | 0.4029 | 0.2515 | 0.5015 |
| **GMMEE-CKF** | **0.1254** | **0.0243** | **0.1558** |

## 4.8 *TSGA-based kernel optimization*

### 4.8.1 *Performance evaluation of TSGA-based kernel selection*

To validate the effectiveness of the proposed TSGA-based kernel selection strategy for the GMMEE-SRCKF algorithm, two complementary experiments have been conducted.

Firstly, a comparative analysis was conducted against manually selected kernels under various operating conditions. The kernels optimized by TSGA consistently yield lower estimation errors. As demonstrated in Table 9, under the UDDS 25°C condition, the MAE is significantly reduced from 0.2803 to 0.0411, and the MSE from 0.0842 to 0.0027. Even in challenging environments, such as US06 at −10°C, TSGA-selected kernels outperform the manually chosen counterparts, demonstrating not only higher accuracy but also greater robustness.

**Table 9**
Performance comparison between TSGA-selected and manually selected kernels.

|  | Manual | | | TSGA | | |
|---|---|---|---|---|---|---|
|  | MAE | MSE | RMSE | MAE | MSE | RMSE |
| UDDS 40°C | 0.4654 | 0.2915 | 0.5399 | 0.3171 | 0.1226 | 0.3501 |
| UDDS 25°C | 0.2803 | 0.0842 | 0.2902 | 0.0411 | 0.0027 | 0.0517 |
| UDDS -10°C | 0.541 | 0.355 | 0.5958 | 0.3234 | 0.1394 | 0.3733 |
| US06 40°C | 0.2455 | 0.0736 | 0.2713 | 0.1537 | 0.032 | 0.1789 |
| US06 25°C | 0.1734 | 0.0384 | 0.1960 | 0.1159 | 0.0193 | 0.1389 |
| US06 -10°C | 0.3306 | 0.1568 | 0.3960 | 0.1920 | 0.0490 | 0.2214 |

Secondly, to further confirm the robustness of the optimization results, 100 independent Monte Carlo simulations are carried out on the LiFePO₄ battery under the WLTP operating condition at 50 °C, considering Uniformly mixed measurement noise. As illustrated in Fig. 13, the manually selected kernel parameters are set to $\eta = 0.5$, $\alpha_1 = 3$, $\alpha_2 = 2.4$, $\beta_1 = 0.009$, $\beta_2 = 0.00005$, while the kernel parameters optimized by the proposed TSGA remain consistent with the settings in Section 4.7. The RMSE results clearly confirm that TSGA-based kernel selection achieves consistently lower errors than manual selection.

These results demonstrate that the TSGA optimization strategy not only improves the estimation accuracy of the GMMEE-SRCKF but also enhances its robustness across varying temperature and driving conditions, effectively removing the reliance on manual kernel parameter adjustment.

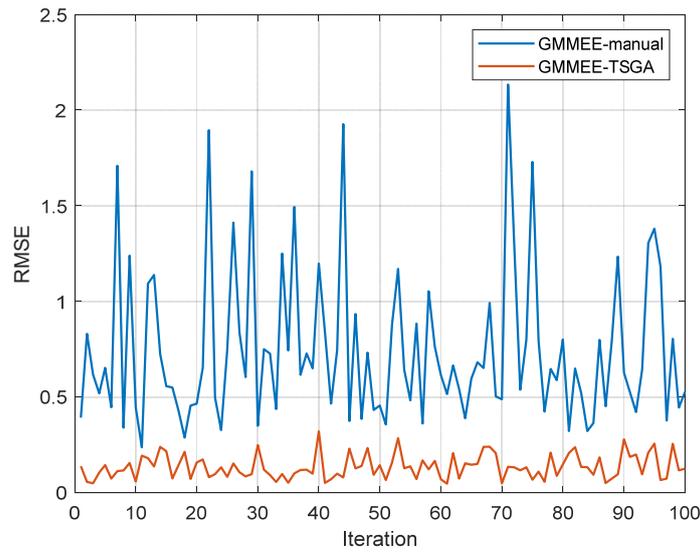

**Fig. 13** RMSE of Monte Carlo simulations.

*4.8.2 The analysis of control parameters of TSGA*

In this section, the effects of three key control parameters of the TSGA are analyzed under condition of Uniformly mixed noise. In each case, one parameter is varied while the others are held constant. As shown in Fig. 14, increasing the population size improves convergence stability and reduces the RMSE of SOC estimation, primarily due to enhanced population diversity. However, larger populations also require greater computational resources. Regarding the search tendency parameter ST, a smaller value may introduce excessive perturbations via the GA, reducing solution stability, whereas a larger value reinforces local search. When ST is appropriately chosen, the incorporation of GA helps maintain moderate population diversity, thereby facilitating the discovery of superior solutions. Finally, a smaller inertia factor accelerates the convergence of seeds toward the global optimum, while an excessively large inertia factor can lead to overshooting, thus impairing

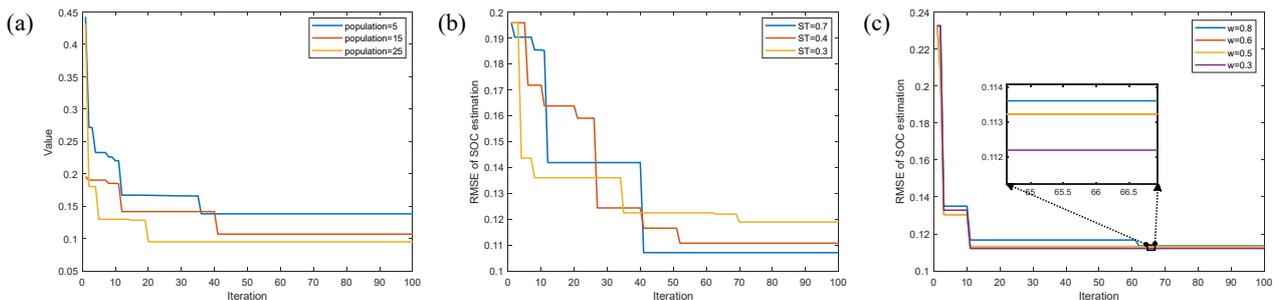

**Fig. 14** Analysis results of the control parameters of TSGA: (a) population size (b) search tendency (c) inertia weight.

both convergence accuracy and speed.

## 5 Conclusion

In this paper, a novel GMMEE-SRCKF is presented as a solution to the complex non-Gaussian noise encountered in the SOC estimation. The employment of two flexible generalized kernel functions in GMMEE provides a more effective approach to fitting non-Gaussian noise. Experiments across varying working conditions verify its robustness, and computational analysis shows high accuracy with minimal time overhead. To automate and optimize kernel selection, a tree seed genetic algorithm (TSGA) is introduced. The TSGA-based strategy consistently outperforms manual selection, significantly reducing estimation errors under diverse conditions. The integration of TSGA with GMMEE-SRCKF results in a more accurate, adaptive SOC estimation framework, demonstrating clear advantages over existing robust algorithms.

While this study proposes a reliable SOC estimation method for lithium-ion batteries with parameter updates achieved via an optimization algorithm, the unknown system model in complex dynamics operating environments is not fully considered. Therefore, the future research will focus on SOC estimation considering battery model uncertainties, with the aim of further improving the robustness and adaptability of the proposed method for lithium-ion batteries in electric vehicles.

## APPENDIX

This appendix provides the full derivation of Eqs. (31) and (32), which were outlined in Section 3.1. By setting the gradient of the cost function $J_L(\mathbf{x}_\tau)$ regarding $\mathbf{x}_\tau$ to zero, we can obtain:

$$\frac{\partial J_L(\mathbf{x}_\tau)}{\partial \mathbf{x}_\tau} = l_1 (\mathbf{W}_\tau)^{\mathrm{T}} (\mathbf{\Delta}_\tau^{\alpha_1,\beta_1} - \mathbf{\Lambda}_\tau^{\alpha_1,\beta_1}) \mathbf{e}_\tau \\ + l_2 (\mathbf{W}_\tau)^{\mathrm{T}} (\mathbf{\Delta}_\tau^{\alpha_2,\beta_2} - \mathbf{\Lambda}_\tau^{\alpha_2,\beta_2}) \mathbf{e}_\tau = 0 \tag{37}$$

with

$$l_1 = \eta \alpha_1 / \beta_1^{\alpha_1}, l_2 = (1-\eta)\alpha_2 / \beta_2^{\alpha_2}$$

$$\mathbf{\Lambda}_{\tau,\mathbf{ij}}^{\alpha_1,\beta_1} = G_{\alpha_1,\beta_1}(e_j(\tau)-e_i(\tau)) |e_j(\tau)-e_i(\tau)|^{\alpha_1-2}$$

$$\mathbf{\Lambda}_{\tau,\mathbf{ij}}^{\alpha_2,\beta_2} = G_{\alpha_1,\beta_1}(e_j(\tau)-e_i(\tau)) |e_j(\tau)-e_i(\tau)|^{\alpha_2-2}$$

$$\Delta_\tau^{\alpha_c,\beta_c} = \text{diag} \left\{ \begin{array}{l} \sum_{i=1}^{L} G_{\alpha_1,\beta_1}(e_1(\tau)-e_i(\tau))|e_1(\tau)-e_i(\tau)|^{\alpha_c-2}, \\ \ldots, \sum_{i=1}^{L} G_{\alpha_1,\beta_1}(e_N(\tau)-e_i(\tau))|e_L(\tau)-e_i(\tau)|^{\alpha_c-2} \end{array} \right\} (c=1,2)$$

From Eq. (37), $\mathbf{x}_\tau$ can be calculated by a fixed-point iteration.

$$\hat{\mathbf{x}}_\tau = g(\hat{\mathbf{x}}_\tau) = \left( l_1 \mathbf{W}_\tau^T \mathbf{\Omega}^{\alpha_1,\beta_1} \mathbf{W}_\tau + l_2 \mathbf{W}_\tau^T \mathbf{\Omega}^{\alpha_2,\beta_2} \mathbf{W}_\tau \right)^{-1} \\ \times \left( l_1 \mathbf{W}_\tau^T \mathbf{\Omega}^{\alpha_1,\beta_1} \mathbf{D}_\tau + l_2 \mathbf{W}_\tau^T \mathbf{\Omega}^{\alpha_2,\beta_2} \mathbf{D}_\tau \right) \tag{38}$$

with

$$\mathbf{\Omega}^{\alpha_c,\beta_c} = \left(\Delta_\tau^{\alpha_c,\beta_c}\right)^T \Delta_\tau^{\alpha_c,\beta_c} + \left(\Lambda_\tau^{\alpha_c,\beta_c}\right)^T \Lambda_\tau^{\alpha_c,\beta_c} = \begin{bmatrix} \mathbf{\Omega}_{xx}^{\alpha_c,\beta_c} & \mathbf{\Omega}_{zx}^{\alpha_c,\beta_c} \\ \mathbf{\Omega}_{xz}^{\alpha_c,\beta_c} & \mathbf{\Omega}_{zz}^{\alpha_c,\beta_c} \end{bmatrix} (c=1,2)$$

According to Eqs. (25), (26), and (38), we arrive at:

$$l_c (\mathbf{W}_\tau)^T \mathbf{\Omega}^{\alpha_c,\beta_c} \mathbf{W}_\tau = l_c \begin{bmatrix} (\mathbf{B}_{p,\tau|\tau-1}^{-1})^T \mathbf{\Omega}_{xx}^{\alpha_c,\beta_c} + \\ \mathbf{\bar{H}}_\tau^T (\mathbf{B}_{r,\tau}^{-1})^T \mathbf{\Omega}_{xz}^{\alpha_c,\beta_c} \end{bmatrix} (\mathbf{B}_{p,\tau|\tau-1}^{-1}) \\ + l_c \left[ (\mathbf{B}_{p,\tau|\tau-1}^{-1})^T \mathbf{\Omega}_{zx}^{\alpha_c,\beta_c} + \mathbf{\bar{H}}_\tau^T (\mathbf{B}_{r,\tau}^{-1})^T \mathbf{\Omega}_{zz}^{\alpha_c,\beta_c} \right] (\mathbf{B}_{r,\tau}^{-1}) \mathbf{\bar{H}}_\tau^T \\ l_c (\mathbf{W}_\tau)^T \mathbf{\Omega}^{\alpha_c,\beta_c} \mathbf{D}_\tau = l_c \begin{bmatrix} (\mathbf{B}_{p,\tau|\tau-1}^{-1})^T \mathbf{\Omega}_{xx}^{\alpha_c,\beta_c} + \\ \mathbf{\bar{H}}_\tau^T (\mathbf{B}_{r,\tau}^{-1})^T \mathbf{\Omega}_{xz}^{\alpha_c,\beta_c} \end{bmatrix} (\mathbf{B}_{p,\tau|\tau-1}^{-1}) \hat{\mathbf{x}}_{\tau|\tau-1} \\ + l_c \left[ (\mathbf{B}_{p,\tau|\tau-1}^{-1})^T \mathbf{\Omega}_{zx}^{\alpha_c,\beta_c} + \mathbf{\bar{H}}_\tau^T (\mathbf{B}_{r,\tau}^{-1})^T \mathbf{\Omega}_{zz}^{\alpha_c,\beta_c} \right] (\mathbf{B}_{r,\tau}^{-1}) \mathbf{y}_\tau (c=1,2) \tag{39}$$

Following Eq. (39), Eq. (38) can be rewritten as:

$$\hat{\mathbf{x}}_\tau = \left( (l_1 \Phi_1 + l_2 \Phi_2) + (l_1 \Phi_3 + l_2 \Phi_4) \Phi_5 \right)^{-1} \\ \times \left( (l_1 \Phi_1 + l_2 \Phi_2) \hat{\mathbf{x}}_{\tau|\tau-1} + (l_1 \Phi_3 + l_2 \Phi_4) \mathbf{y}_\tau \right) \tag{40}$$

with

$$\Phi_1 = \left[ (\mathbf{B}_{p,\tau|\tau-1}^{-1})^T \mathbf{\Omega}_{xx}^{\alpha_1,\beta_1} + \mathbf{\bar{H}}_\tau^T (\mathbf{B}_{r,\tau}^{-1})^T \mathbf{\Omega}_{xz}^{\alpha_1,\beta_1} \right] \mathbf{B}_{p,\tau|\tau-1}^{-1}$$

$$\Phi_2 = \left[ (\mathbf{B}_{p,\tau|\tau-1}^{-1})^T \mathbf{\Omega}_{xx}^{\alpha_2,\beta_2} + \mathbf{\bar{H}}_\tau^T (\mathbf{B}_{r,\tau}^{-1})^T \mathbf{\Omega}_{xz}^{\alpha_2,\beta_2} \right] \mathbf{B}_{p,\tau|\tau-1}^{-1}$$

$$\Phi_3 = \left[ (\mathbf{B}_{p,\tau|\tau-1}^{-1})^T \mathbf{\Omega}_{zx}^{\alpha_1,\beta_1} + \mathbf{\bar{H}}_\tau^T (\mathbf{B}_{r,\tau}^{-1})^T \mathbf{\Omega}_{zz}^{\alpha_1,\beta_1} \right] \mathbf{B}_{r,\tau}^{-1}$$

$$\Phi_4 = \left[ (\mathbf{B}_{p,\tau|\tau-1}^{-1})^T \mathbf{\Omega}_{zx}^{\alpha_2,\beta_2} + \mathbf{\bar{H}}_\tau^T (\mathbf{B}_{r,\tau}^{-1})^T \mathbf{\Omega}_{zz}^{\alpha_2,\beta_2} \right] \mathbf{B}_{r,\tau}^{-1}$$

$$\Phi_5 = \mathbf{\bar{H}}_\tau^T$$

By using the matrix inversion lemma, Eq. (40) can be rewritten as Eq. (31), where $\hat{\mathbf{K}}_\tau$ can be obtained by Eq. (41), which can be used to calculate the posterior covariance matrix Eq. (32):

$$\hat{\mathbf{K}}_\tau = \begin{bmatrix} l_1 \tilde{\mathbf{P}}_{\tau|\tau-1,xx}^{\alpha_1,\beta_1} + l_2 \tilde{\mathbf{P}}_{\tau|\tau-1,xx}^{\alpha_2,\beta_2} + \mathbf{\bar{H}}_\tau^T (l_1 \tilde{\mathbf{P}}_{\tau|\tau-1,xz}^{\alpha_1,\beta_1} + l_2 \tilde{\mathbf{P}}_{\tau|\tau-1,xz}^{\alpha_2,\beta_2}) \\ + [(l_1 \tilde{\mathbf{P}}_{\tau|\tau-1,zx}^{\alpha_1,\beta_1} + l_2 \tilde{\mathbf{P}}_{\tau|\tau-1,zx}^{\alpha_2,\beta_2}) + \mathbf{\bar{H}}_\tau^T (l_1 \tilde{\mathbf{P}}_{\tau|\tau-1,zz}^{\alpha_1,\beta_1} + l_2 \tilde{\mathbf{P}}_{\tau|\tau-1,zz}^{\alpha_2,\beta_2})] \mathbf{\bar{H}}_\tau^T \end{bmatrix}^{-1} \\ \times [(l_1 \tilde{\mathbf{P}}_{\tau|\tau-1,zx}^{\alpha_1,\beta_1} + l_2 \tilde{\mathbf{P}}_{\tau|\tau-1,zx}^{\alpha_2,\beta_2}) + \mathbf{\bar{H}}_\tau^T (l_1 \tilde{\mathbf{P}}_{\tau|\tau-1,zz}^{\alpha_1,\beta_1} + l_2 \tilde{\mathbf{P}}_{\tau|\tau-1,zz}^{\alpha_2,\beta_2})] \tag{41}$$

with

$$\tilde{\mathbf{P}}^{\alpha_c,\beta_c}_{\tau|\tau-1,xx} = \left(\mathbf{B}_{p,\tau|\tau-1}^{-1}\right)^T \mathbf{\Omega}^{\alpha_c,\beta_c}_{xx} \mathbf{B}_{p,\tau|\tau-1}^{-1}$$

$$\tilde{\mathbf{P}}^{\alpha_c,\beta_c}_{\tau|\tau-1,xz} = \left(\mathbf{B}_{r,\tau}^{-1}\right)^T \mathbf{\Omega}^{\alpha_c,\beta_c}_{xz} \mathbf{B}_{p,\tau|\tau-1}^{-1}$$

$$\tilde{\mathbf{P}}^{\alpha_c,\beta_c}_{\tau|\tau-1,zx} = \left(\mathbf{B}_{p,\tau|\tau-1}^{-1}\right)^T \mathbf{\Omega}^{\alpha_c,\beta_c}_{zx} \mathbf{B}_{r,\tau}^{-1}$$

$$\tilde{\mathbf{P}}^{\alpha_c,\beta_c}_{\tau|\tau-1,zz} = \left(\mathbf{B}_{r,\tau}^{-1}\right)^T \mathbf{\Omega}^{\alpha_c,\beta_c}_{zz} \mathbf{B}_{r,\tau}^{-1}$$